\documentclass{PoS}

\title{New measurement and QCD analysis of DIS data from HERA}

\ShortTitle{New measurement and QCD analysis of DIS data from HERA}

\author{\speaker{Zhiqing Zhang}\thanks{For the H1 Collaboration.}\\
        Laboratoire de l'Acc\'el\'erateur Lin\'eaire, Univ.\ Paris-Sud 11 et IN2P3/CNRS, France\\
        E-mail: \email{zhang@lal.in2p3.fr}}

\abstract{This talk covers three contributions from H1: ``Measurement of the inclusive $e^{\pm}p$ scattering cross section at high inelasticity $y$ and of the structure function $F_L$'', ``Determination of the integrated luminosity at HERA using elastic QED Compton events" and ``Inclusive deep inelastic scattering at high $Q^2$ with longitudinally polarized lepton beams at HERA". These are new measurements mainly based on the full HREA-II data but include also those from HERA-I in the combination whenever it is relevant. The main results of these measurements are briefly summarized here.}

\FullConference{36th International Conference on High Energy Physics,\\
		July 4-11, 2012\\
		Melbourne, Australia}

\begin{document}

\section{Introduction}
The $ep$ collider HERA used to be the largest electron microscope of the world. The inclusive neutral current (NC) and charged current (CC) deep inelastic scattering (DIS) cross section data measured by H1 and ZEUS at HERA-I (1992-2000)~\cite{h1zeus-hera1}, where the unpolarized electron\footnote{In these proceedings, ``electron" refers generically to both electrons and positrons. Where distinction is required the terms $e^-$ and $e^+$ are used.} beam of up to 27.5\,GeV collided with the unpolarized proton beam of up to 920\,GeV, have been the primary source for constraining of parton distribution functions (PDFs) of the proton. In HERA-II (2003-2007), spin rotators were installed around the interaction points of H1 and ZEUS to provide longitudinally polarized electron beam there. Most of the data at HERA-II were taken at the nominal proton beam energy $E_p$ of 920\,GeV in four distinct data sets: $e_R^-p$, $e_L^-p$, $e_R^+p$ and $e_R^+p$, i.e.\ right- and left-handed polarized electron- and positron beams scattering with proton beam. The corresponding integrated luminosity (mean polarization $P_e$) values are 47.3\,pb$^{-1}$  ($+36.0$\%), 104.4\,pb$^{-1}$ ($-25.8$\%), 101.3\,pb$^{-1}$ ($+32.5$\%) and 80.7\,pb$^{-1}$ ($-37.0$\%), respectively. The $e^-p$ data corresponds to an almost tenfold increase in luminosity over the HERA-I data set. Towards the end of data taking, two data sets with lower proton beam energies of 575\,GeV and 460\,GeV were taken for a direct measurement of longitudinal structure function $F_L$ (Sec.~\ref{sec:fl})~\cite{fl}.

For a precise cross section measurement, one important input is the integrated luminosity value, which has recently been determined for HERA-II data using elastic QED Compton events
(Sec.~\ref{sec:lumi})~\cite{lumi}.

The other new results just released before the conference concern the final NC and CC cross sections at high $Q^2$, four momentum transfer squared, measured using all HEAR-II data and a new QCD analysis of all H1 data to obtain a new set of PDFs, H1PDF\,2012 (Sec.~\ref{sec:hiq2})~\cite{hiq2-hera2}.
 
To appreciate the constraint on PDFs from the inclusive NC and CC cross sections, it is helpful to write down explicitly the expressions of the differential NC and CC cross sections in terms of different structure functions  
\begin{eqnarray}
\frac{d^2\sigma^\pm_{\rm NC}}{dxdQ^2}&=&\frac{2\pi\alpha^2}{xQ^4}\left(Y_+\tilde{F}^\pm_2\mp Y_-x\tilde{F}_3^\pm -y^2\tilde{F}_L^\pm\right)\,,\\
\frac{d^2\sigma^\pm_{\rm CC}}{dxdQ^2}&=&(1\pm P_e)\frac{G^2_F}{2\pi x}\left[\frac{M^2_W}{M^2_W+Q^2}\right]^2\left(Y_+W_2^\pm\mp Y_-xW_3^\pm-y^2W_L^\pm\right)\,,
\end{eqnarray}
where $Y_\pm=1\pm(1-y)^2$ and $y$ characterizes the inelasticity of the interaction and is related to Bjorken $x$, $Q^2$ and center-of-mass energy $\sqrt{s}$ by $y=Q^2/(xs)$, $\alpha\equiv\alpha(Q^2=0)$ is the fine structure constant, $G_F$ is the Fermi constant and $M_W$ is the propagator mass of the $W$ boson. For the NC interaction, the generalized structure functions $\tilde{F}_2^\pm$ and $x\tilde{F}_3^\pm$ can be further decomposed as
\begin{eqnarray}
\tilde{F}_2^\pm&=&F_2-(v_e\pm P_ea_e)\kappa\frac{Q^2}{Q^2+M^2_Z}F_2^{\gamma Z}+(a^2_e+v^2_e\pm P_e2v_ea_e)\kappa^2\left[\frac{Q^2}{Q^2+M^2_Z}\right]^2F_2^Z\,,\label{eq:f2}\\
x\tilde{F}_3^\pm&=&-(a_e\pm P_ev_e)\kappa\frac{Q^2}{Q^2+M^2_Z}xF_3^{\gamma Z}+(2a_ev_e\pm P_e[v^2_e+a^2_e])\kappa^2\left[\frac{Q^2}{Q^2+M^2_Z}\right]^2xF_3^Z\,,
\end{eqnarray}
with $M_Z$ being the mass of the $Z$ boson, $\kappa^{-1}=4\frac{M^2_W}{M^2_Z}\left(1-\frac{M^2_W}{M^2_Z}\right)$ in the on-mass-shell scheme, $v_e$ and $a_e$ the vector and axial-vector couplings of the electron to the $Z$ boson. In Eq.(\ref{eq:f2}), $F_2$ corresponds to the dominant electromagnetic structure function of photon exchange, the other structure function terms are due to photon-$Z$ interference and pure $Z$ exchange. The generalized longitudinal structure function $\tilde{F}_L$ may be similarly decomposed. 

In the quark-parton model, the structure functions are related to linear combinations of quark and anti-quark momentum distributions $xq(x,Q^2)$ and $x\bar{q}(x,Q^2)$ and $F_L$ is zero because of helicity conservation. In QCD, the gluon emission gives rise to a non-vanishing $F_L$. Measuring the structure function $F_L$ therefore provides a way of studying the gluon density and a test of perturbative QCD.

\section{Measurement of longitudinal structure function $F_L(x,Q^2)$}\label{sec:fl}
Using DIS events with the scattered electron detected in the backward lead-scintillator calorimeter from several independent data samples taken by the H1 detector from HERA-II, inclusive NC DIS cross sections have been measured. The cross section measurement at the nominal $E_p=920$\,GeV covers the $Q^2$ range from 2.5\,GeV$^2$ to 90\,GeV$^2$. The cross section measurement at reduced proton beam energies, $E_p=575$\,GeV and $E_p=460$\,GeV, covers the similar kinematic domain of $1.5\leq Q^2\leq 90\,{\rm GeV}^2$ and uses integrated luminosity of $5.9\,{\rm pb}^{-1}$ and $12.2\,{\rm pb}^{-1}$, respectively. The data at $E_p=920$\,GeV significantly improve the accuracy of the cross section measurements at high $y$ when compared to the previous H1 data. All data are combined with the HERA-I results to provide a new accurate data sample covering $0.2\leq Q^2\leq 150\,{\rm GeV}^2$, $5\times 10^{-6}<x<0.15$ and $0.005<y<0.85$ with a total uncertainty of 1\% at low $Q^2$ and low $x$.

\begin{figure}[htb]
\begin{center}
\includegraphics[width=.6\textwidth]{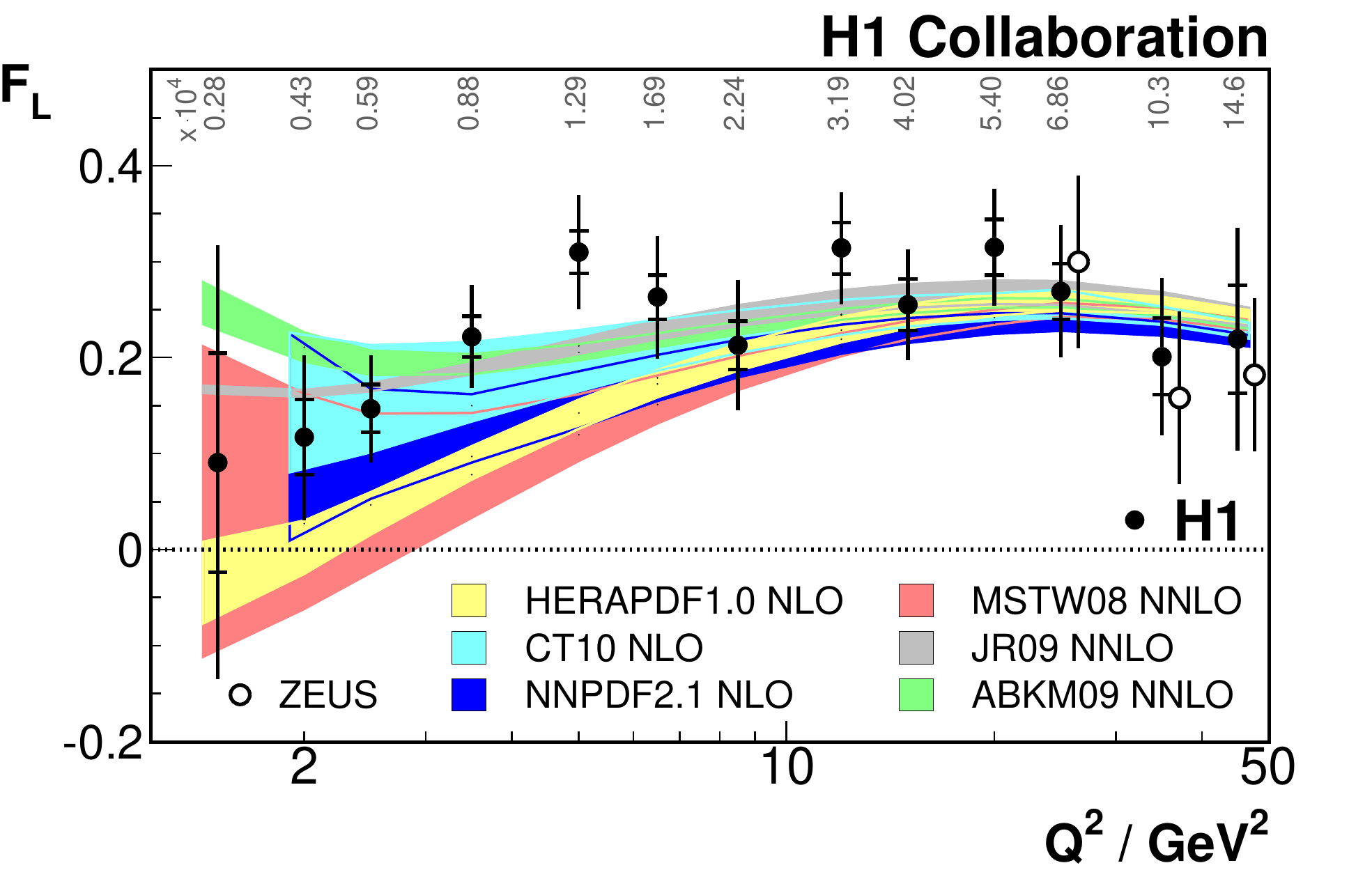}
\end{center}
\vspace{-7mm}
\caption{Proton structure function $F_L$ shown as a function of $Q^2$. The average $x$ values for each $Q^2$ are indicated. The inner and outer error bars represent statistical and total errors, respectively. The bands represent predictions based on HERAPDF1.0, CT10 and NNPDF2.1 NLO as well as MSTW08, GJR08 and ABKM09 NNLO calculations.}
\label{fl}
\end{figure}

The data at $E_p=460$\,GeV and $E_p=575$\,GeV, together with the measurements at $E_p=920$\,GeV are used to determine the structure function $F_L$. The measurement covering $1.5\leq Q^2\leq 45\,{\rm GeV}^2$ and $2.7\times 10^{-5}<x<2\times 10^{-3}$ extends previous measurements to lower $Q^2$ and $x$. The data are reasonably well reproduced by the predictions based on NLO and NNLO QCD (Fig.\ref{fl}).


\section{Determination of integrated luminosity using elastic QED Compton events}\label{sec:lumi}
At HERA, the integrated luminosity is usually measured using dedicated detectors located at small angles to detect photon and/or electron emitted almost collinearly to the incident electron in the Bethe-Heitler (BH) process $ep\to e\gamma p$. The advantage of this process is its large cross section, thus the statistical uncertainty of the corresponding luminosity measurement is negligible. However it suffers from a few sizable systematic uncertainties associated e.g.\ to the acceptance limitations of the small angle detectors and to satellite corrections for those $ep$ collisions happening outside the nominal interaction region.

A new determination of the integrated luminosity is realized using the elastic QED Compton events with the final state electron and photon detected and measured in the main detector. This method is thus insensitive to details of the beam optics. However, the smallness of the cross section leads to limited statistical precision for any given small amount of integrated luminosity. Therefore all data collected from HERA-II are analyzed together resulting in a measurement with a statistical uncertainty of 0.8\% and a systematic error of 2.1\%. The measurement is in agreement with the BH measurement which has a total uncertainty of over 3.0\%. 

\section{New measurement of DIS data at high $Q^2$ and QCD analysis all H1 data}\label{sec:hiq2}
Using the four HREA-II data sets mentioned in the introduction, the total CC $\sigma^{\rm tot}_{\rm CC}$, single differential CC and NC $d\sigma_{\rm CC, NC}/dQ^2$ and double differential CC and NC $d^2\sigma_{\rm CC, NC }/dxdQ^2$ cross sections are measured~\cite{hiq2-hera2}.

\begin{figure}[htb]
\begin{center}
\includegraphics[width=.475\textwidth]{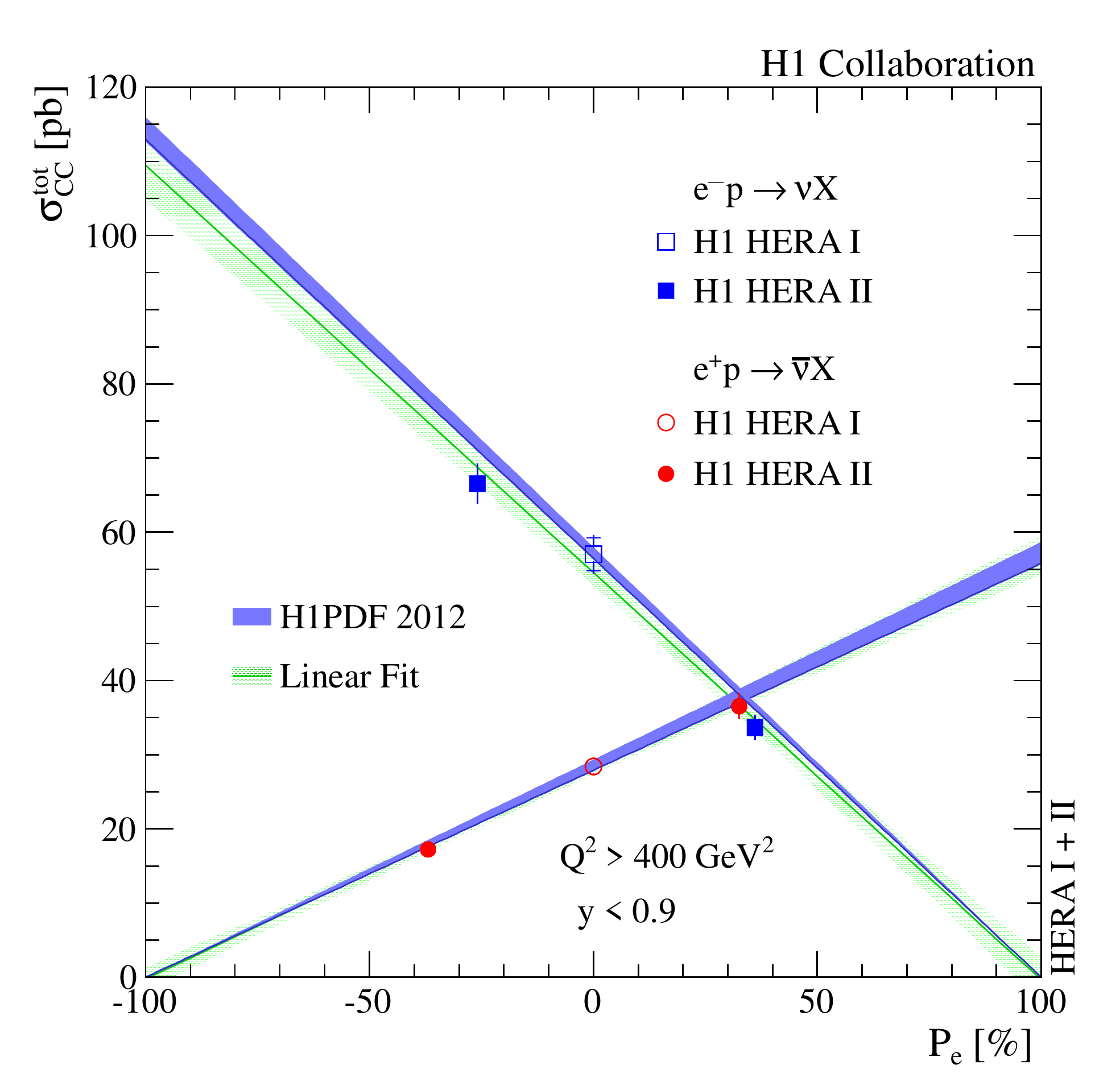}
\includegraphics[width=.475\textwidth]{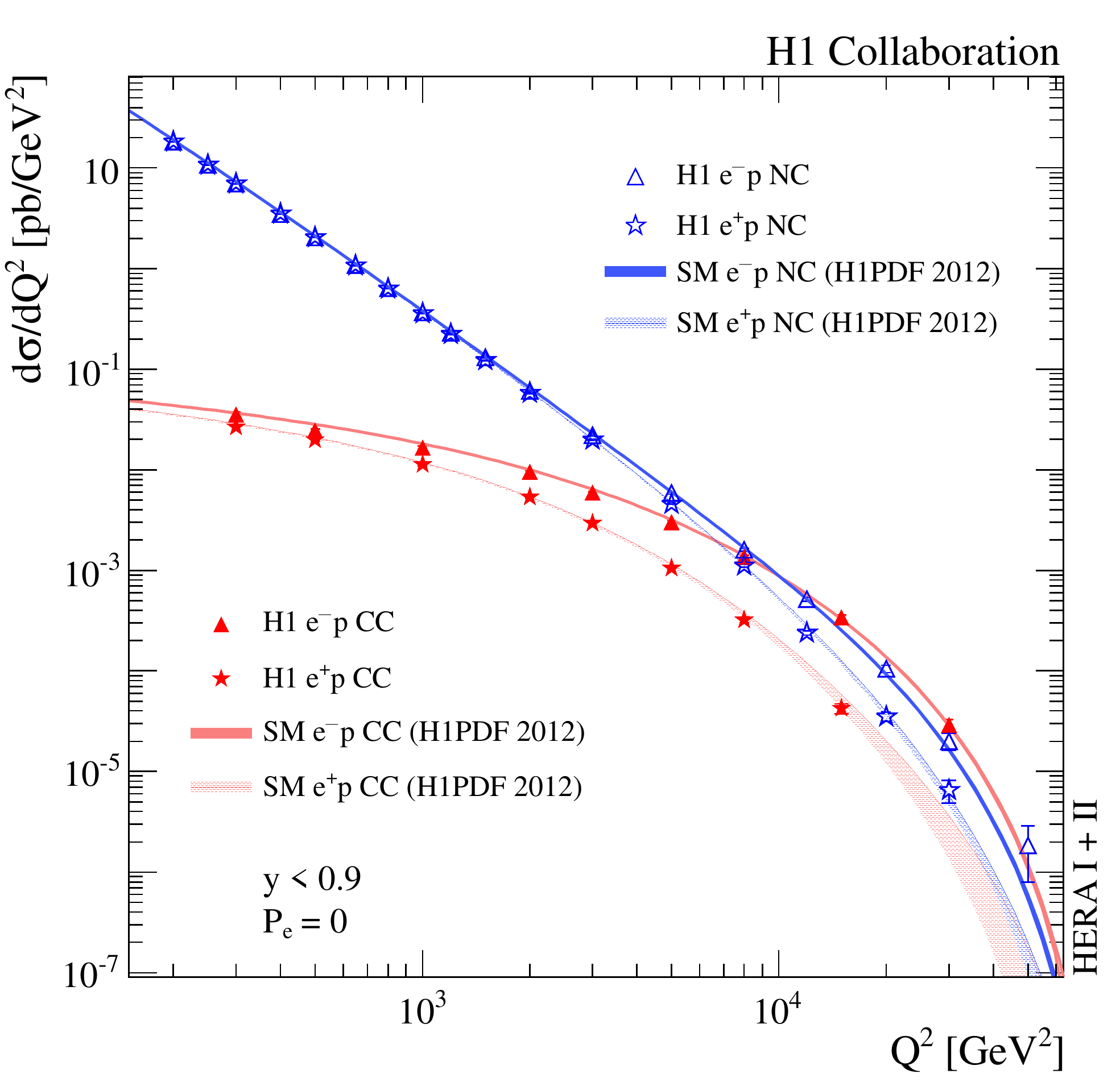}
\end{center}
\vspace{-5mm}
\caption{\underline{Left}: Dependence of the $e^\pm p$ CC cross sections on the longitudinal lepton beam polarization $P_e$. \underline{Right}: $Q^2$ dependence of the NC and CC cross sections $d\sigma/dQ^2$ for the combined HERA-I+II unpolarized $e^\pm p$ data. In both figures, the inner and outer error bars represent the statistical and total errors, respectively.}
\label{fig:cctot-nccc}
\end{figure}

The total CC cross sections $\sigma^{\rm tot}_{\rm CC}$ for $e_{L, R}^\pm p$ scatterings are measured in the kinematical region of $Q^2>400\,{\rm GeV}^2$ and $y<0.9$. Together with the corresponding cross sections measured at HERA-I with unpolarized $e^\pm$ beams, the linear dependence of the cross sections on $P_e$ (Fig.\,\ref{fig:cctot-nccc}(left)) can be verified for both $e^-p$ and $e^+p$ interactions (in the latter case with more data than that previously published~\cite{cctot05}). The data are consistent with the absence of right handed weak charged currents. 

The single differential NC and CC cross sections $d\sigma_{\rm NC,CC}/dQ^2$ are measured for $y<0.9$ and in $Q^2$ varying over more than two orders of magnitude (the combined version with unpolarized HERA-I $e^\pm p$ data is shown in Fig.\,\ref{fig:cctot-nccc}(right)). The NC cross sections exceed the CC cross sections at $Q^2\simeq 200\,{\rm GeV}^2$ by more than two orders of magnitude. The steep decrease of the NC cross section with increasing $Q^2$ is due to the dominating photon exchange cross section which is proportional to $1/Q^4$. In contrast the CC cross section is proportional to $[M_W^2 /(Q^2 + M_W^2 )]^2$ and approaches a constant value at $Q^2 \simeq 300\,{\rm GeV}^2$. The NC and CC cross sections are of comparable size at $Q^2 \sim 10^4\,{\rm GeV}^2$, where the photon and $Z$ exchange contributions to the NC process are of similar size to those of $W^\pm$ exchange to the CC process. These measurements thus illustrate the unified behavior of the electromagnetic and the weak interactions in DIS.

The double differential NC and CC cross sections $d^2\sigma_{\rm NC, CC}$ are also measured in the kinematic range $120\leq Q^2\leq 50\,000\,{\rm GeV}^2$ and $0.002\leq x\leq 0.65$ for NC, and $300\leq Q^2\leq 30\,000\,{\rm GeV}^2$ and $0.008\leq x\leq 0.4$ for CC. Both the statistical and systematic precisions have substantially improved with respect to the corresponding measurements from HERA-I with the unpolarized electron beams. There is a special measurement at high $y$ region of $0.63<y<0.9$ and $60\leq Q^2\leq 800\,{\rm GeV}^2$, where the left and right handed polarized data sets are combined to measure unpolarized NC cross sections for $e^-p$ and $e^+p$ as in this region the sensitivity to the polarization is small.

From the NC cross section measurements, a number of other results are derived:
(i) {\it NC polarization asymmetry:} The polarized single differential cross sections $d\sigma_{\rm NC}/dQ^2$ are used to construct a polarization asymmetry. 
The magnitude of the measured asymmetry is observed to increase with increasing $Q^2$ and is positive in $e^+p$ and negative in $e^-p$ scattering. The asymmetry provides a direct measure of parity violation in NC DIS.
(ii) {\it A first determination of $F^{\gamma Z}_2$:} Use the double differential polarized NC cross sections, the function function $F_2^{\gamma Z}$ is extracted for the first time. The determination is performed for $Q^2\leq 200\,{\rm GeV}^2$. The results are quoted at $Q^2=1\,500\,{\rm GeV}^2$ as only a weak $Q^2$ dependence is expected.
(iii) {\it Improved measurement of $xF_3^{\gamma Z}$:} Double differential NC cross sections corresponding to $P_e=0$ from HERA-II obtained by adding L and R data sets and correcting for a small residual polarization are combined with those from HERA-I to update the previous measurement of the structure function $xF_3^{\gamma Z}$~\cite{oldxf3a,oldxf3b}.

\begin{figure}[htb]
\begin{center}
\includegraphics[width=.4\textwidth]{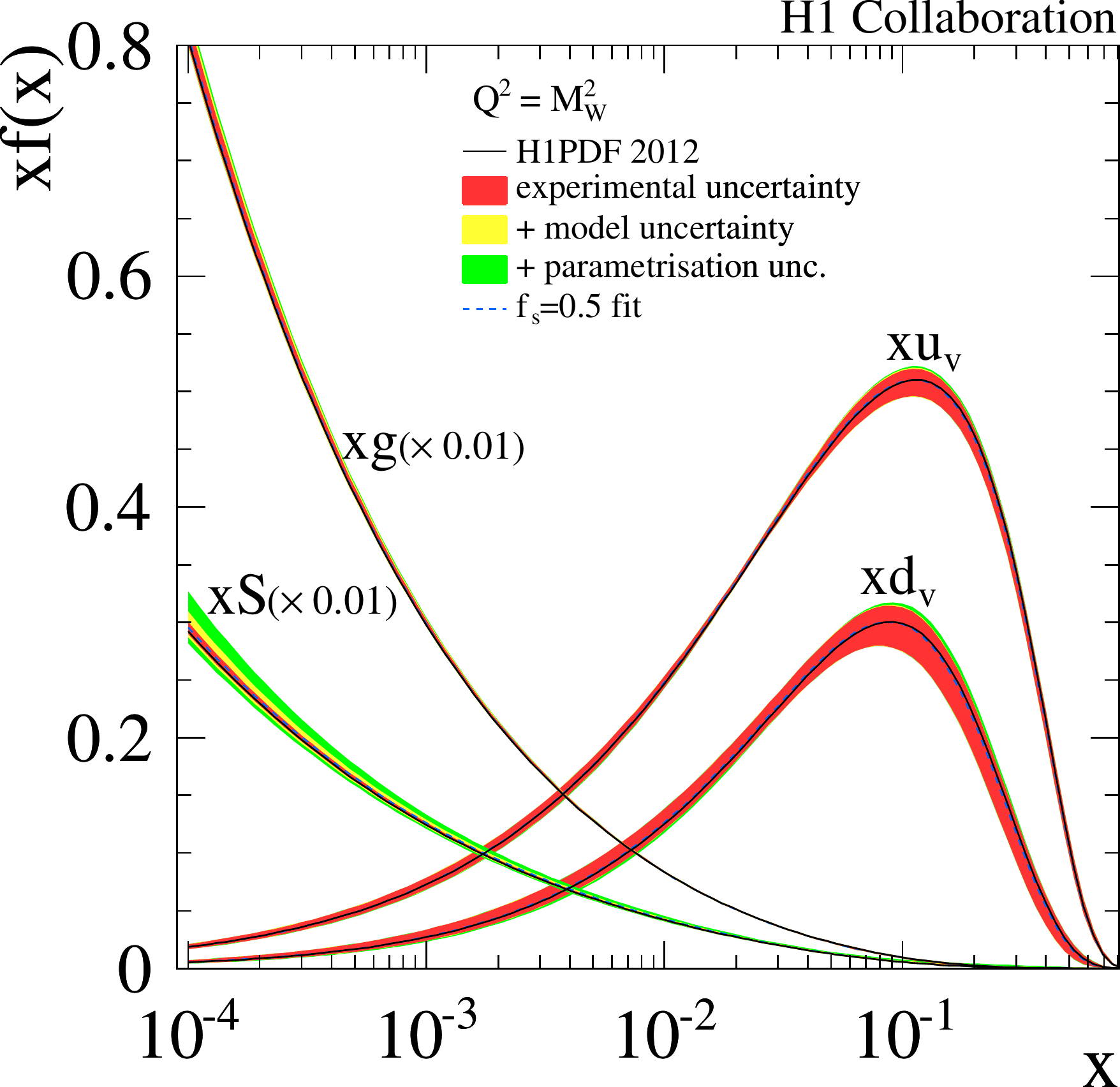}\hspace{2mm}
\includegraphics[width=.58\textwidth]{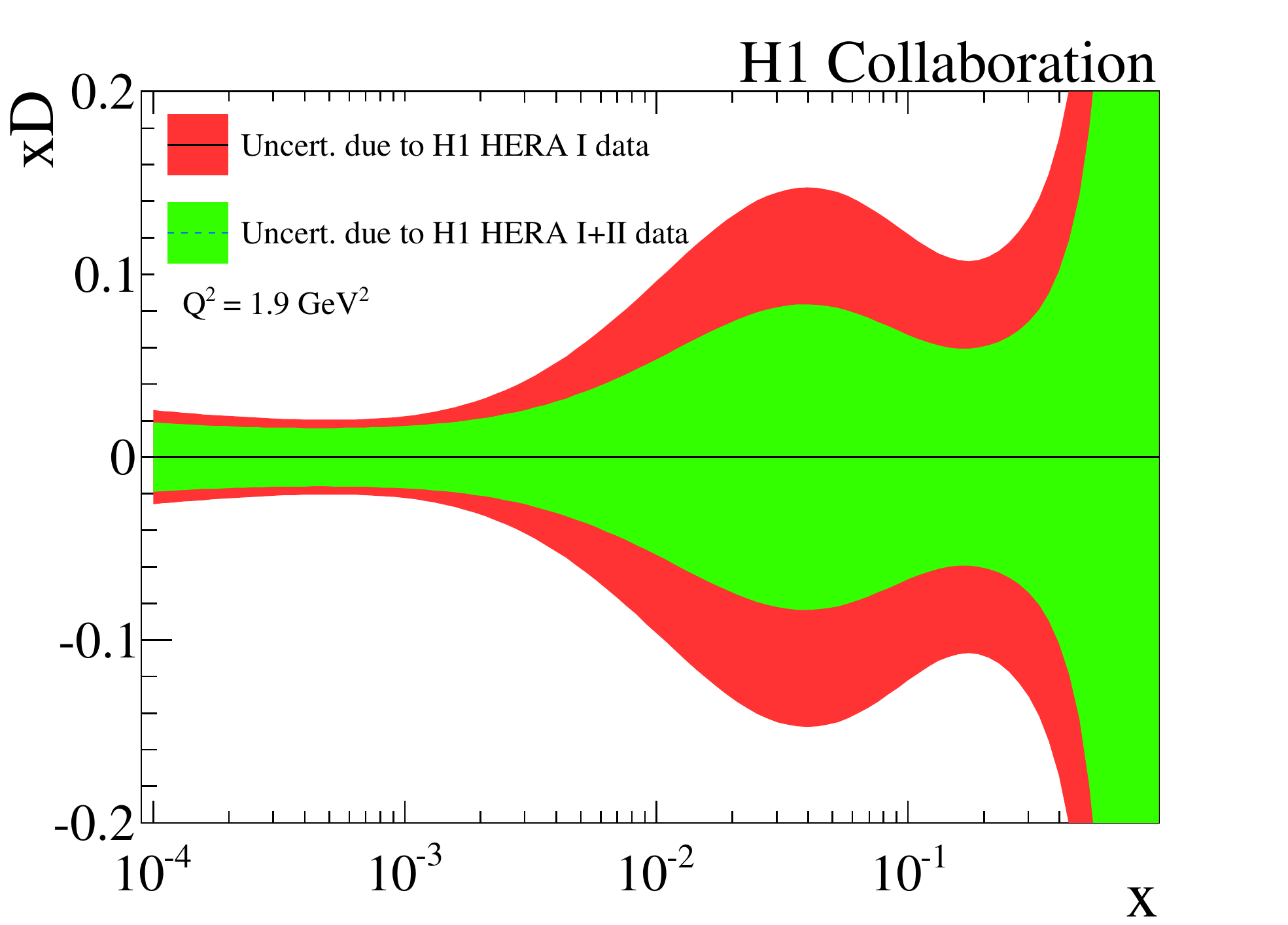}
\end{center}
\vspace{-5mm}
\caption{\underline{Left}: Parton distribution functions of H1PDF\,2012 at the evolved scale of $M^2_W$. The gluon and sea distributions are scaled by a factor 0.01. The uncertainties include the experimental uncertainties (inner), the model uncertainties (middle) and the parametrization variation (outer). All uncertainties are added in quadrature. \underline{Right}: Comparison of relative experimental uncertainties of $xD$ extracted from HERA-I (outer) vs.\ HERA-I+II (inner) data sets under the same fit conditions to better assess the effect of the new high $Q^2$ measurements.}
\label{fig:pdfs}
\end{figure}

To assess the impact of the new H1 NC and CC cross sections at high $Q^2$ measured with the longitudinally polarized electron beams on the determination of PDFs, a new QCD analysis (H1PDF 2012) is performed. In addition to the new HERA-II data, the previously published unpolarized HERA-I data at high $Q^2$ and at low $Q^2$, as well as the H1 measurements at lower proton beam energies are used. This analysis supersedes the previous H1PDF 2009 fit. The resulting PDFs are shown in Fig.\ref{fig:pdfs}(left) and the impact of HERA-II data on the PDF precision is shown as an example for the down quark density $xD$ in Fig.\ref{fig:pdfs}(right). 

\section{Summary}

Using the low $E_p$ data taken at HERA-II in combination with the nominal $E_p$ data from both HERA-I and HERA-II, a direct $F_L$ measurement has been performed at an extended kinematic region $1.5\leq Q^2\leq 45\,{\rm GeV}^2$ and $2.7\times 10^{-5}<x<2\times 10^{-3}$ providing a direct test on the gluon density of the proton at lower $Q^2$ and $x$ than that of previous measurements. The integrated luminosity at HERA-II has been measured with elastic QED Compton events with a total precision of 2.3\% better than the precision obtained with the Bethe-Heitler process for this data set. The NC and CC cross sections at high $Q^2$ with the nominal $E_p$ and all polarized $e^\pm$ data from HERA-II have been finalized. Both the statistical and systematic uncertainties of the measurements have been substantially reduced compared to previous HERA-I H1 data. The new QCD analysis shows a significant impact of the new data on the increased precision of the PDFs. These data are being combined with those from ZEUS and a new QCD fit HERAPDF2.0 is being performed. The combined data are expected to provide an ultimate precision on PDFs.


\begin{thebibliography}{99}
\bibitem{h1zeus-hera1}  F.D.~Aaron et al.\ (H1 and ZEUS Collab.),
\emph{JHEP} {\bf 1001} (2010) 109 [{\tt arXiv:0911.0884}].
\bibitem{fl} F.D.~Aaron et al.\ (H1 Collab.), 
\emph{Eur.\ Phys.\ J.} C{\bf 71} (2011) 1579 [{\tt arXiv:1012.4355}].
\bibitem{lumi} F.D.~Aaron et al.\ (H1 Collab.), 
\emph{Eur.\ Phys.\ J.} C{\bf 72} (2012) 2163 [{\tt arXiv:1205.2448}].
\bibitem{hiq2-hera2} F.D.~Aaron et al.\ (H1 Collab.), 
\emph{JHEP} {\bf 1209} (2012) 061 [{\tt arXiv:1206.7007}].
\bibitem{cctot05} A.~Aktas et al.\ (H1 Collab.), 
\emph{Phys.\ Lett.} B{\bf 634} (2006) 173 [{\tt hep-ex/0512060}].
\bibitem{oldxf3a} C.~Adloff et al.\  (H1 Collab.), 
\emph{Eur.\ Phys.\ J.} C{\bf 19} (2001) 269 [{\tt hep-ex/0012052}].
\bibitem{oldxf3b}C.~Adloff et al.\ (H1 Collab.), 
\emph{Eur.\ Phys.\ J.} C{\bf 30} (2003) 1 [{\tt hep-ex/0304003}]. 

\end{thebibliography}
\end{document}